\documentstyle[preprint,eqsecnum,aps,prb,amstex]{revtex}

\begin{document}
\title{Structural evidence against current-induced destruction of charge-ordering
in manganese oxides}
\author{A.\ Wahl, V.\ Caignaert, S.\ Mercone}
\address{Laboratoire CRISMAT, UMR 6508, ENSICAEN - Universit\'{e} de Caen,\\
6 Boulevard du Mar\'{e}chal Juin, 14050 Caen Cedex, France.}
\author{F.\ Fa\"{u}th}
\address{European Synchrotron Radiation Facility (ESRF)\\
6, Rue Jules Horowitz, F-38043 Grenoble,France}
\date{\today}
\maketitle

\begin{abstract}
High-resolution x-ray scattering, in the presence of an applied current, has
been used for studying the stability of the charge ordered phase of
manganese oxides upon current biasing. We find that the charge ordered
structure is unchanged when a current is flowing in the sample. Such a
result indicates that the non-linear conduction observed in charge ordered
manganites can not be ascribed to a current-induced destabilisation of
charge ordering.
\end{abstract}

\pacs{}

\smallskip Rare earth manganites with general chemical formula $%
R_{1-x}A_xMnO_3$ (where $R$ and $A$ being trivalent rare-earth and divalent
ions, respectively) show a wide variety of electronic and magnetic
properties such as colossal magnetoresistance and charge (CO) or orbital
ordering (OO).\cite{CO,REVIEW,CMR} The physics of the detabilization of the
CO by different types of perturbation has recently been under very active
investigation.\cite{LEE95,YOS95,TOM95,KIR97,FIE98,OGA98,MOR97,GUH00a} For
instance, it is now well established that the CO is stabilized by lattice
distortions ; thus, a perturbation of this distortion by changing the
average-A-site cationic radius can weaken the CO.\cite{MAR99} Besides, it
has been reported that irradiation by X-rays\cite{KIR97} or light and
application of pressure\cite{FIE98,OGA98,MOR97} can also induce a melting of
the CO, leading to a transition from a charge-localized state (CL) to a
charge-delocalized state (CD).\ Application of an electric field is also
known to induce a CL-CD transformation ; this latter feature is always
associated with a strong non-linearity of the tension-current
characteristics (V-I).\cite{GUH00a,ASA97,GUH00b,RAO00,BUD01}

Up to now, the optical and irradiation induced transition are argued to be
the result of classical percolation transport in a non-homogeneous medium ;
however, it is not clear whether the underlying mechanism is the same for
all perturbations, in particular for the electric field induced transition.
In the latter case, this feature is interpreted in terms of current induced
destabilization of the CO-CL state leading to the creation of low
resistivity conducting path along the current flow.\cite
{GUH00a,ASA97,GUH00b,RAO00,BUD01} Recent experimental results do not agree
with such a current-induced CO melting scenario.\cite{MERCONE}\ Indeed,
electrical measurements, performed on the non-charge ordered compounds Pr$%
_{0.8}$Ca$_{0.2}$MnO$_3$, also exhibit a current induced CL-CD transition
associated with a strong non linearity of the V-I characteristics. In such a
case, the destruction of the CO can not be invoked, suggesting that the
mechanism is more complex than the one initially proposed.

In this paper, we present synchrotron X-ray diffraction data for the
charge-ordered manganites Pr$_{0.63}$Ca$_{0.37}$MnO$_3$ and Nd$_{0.6}$Ca$%
_{0.4}$MnO$_3$ polycrystalline samples. Details for synthesis of the sample
are reported elsewhere.\cite{MAI97} Measurements have been performed as a
function of temperature {\it and current} on the diffractometer ID31 at ESRF
(Grenoble, France) using an incident wavelength of 0.413 \AA\ (E=30 keV).
The typical sizes of the samples allow a homogeneous sample cooling together
with suitable beam transmission. Two contacts of In were soldered onto the
section (1x2 mm$^2$) of the sample, leaving an uncovered section L= 3mm. V-I
data were taken with current biasing (Keithley 236).

Our X-rays diffraction patterns, analyzed with the Rietveld method, lead to
an orthorhombic cell with $a\simeq a_p\sqrt{2},$ $b\simeq 2a_p$, $c\simeq a_p%
\sqrt{2}$ in a $PnmA$ space group at room temperature. Previous
neutron-diffraction measurements\cite{YOS95,JIR85,COX98} showed that the CO
structure is characterized by a doubling of the a-axis and the appearance of
additional peaks of the type (h/2 k l). A careful search was made for such
superlattice peaks associated with charge ordering of $Mn^{3+\text{ }}$and $%
Mn^{4+}$ in Pr$_{0.63}$Ca$_{0.37}$MnO$_3$ and Nd$_{0.6}$Ca$_{0.4}$MnO$_3$
below $T_{CO}$ = 235K and 200K, respectively. Figure 1 shows narrow regions
covering the strongest of these peaks for Nd$_{0.6}$Ca$_{0.4}$MnO$_3$ at a
temperature below $T_{CO}$ ($T=100K$, solid line) together with the results
of similar scans for T=295K 
\mbox{$>$}%
$T_{CO}$ (symbols). In the latter case, no superlattice peaks can be
observed whereas, for T= 100 K the superlattice peaks are well defined. High
resolution X-ray scattering measurements in the presence of an applied
direct current would permit to determine unambiguously whether the current
induced CL-CD transition associated with the non-linear V-I characteristics
can be linked to a CO destruction. Transport properties were simultaneously
measured in situ during the experiments. A strong non-linearity i. e a
deviation from the Ohm's law, is observed when the bias current attains a
threshold value for Pr$_{0.63}$Ca$_{0.37}$MnO$_3$ and Nd$_{0.6}$Ca$_{0.4}$MnO%
$_3$ powder samples.\ This non-linearity is even more obvious when $\frac R{%
Rohmic}$ $vs$ $I$ curves are plotted (See figure 2a and 2b). As expected the
resistance is independent of the bias current in the ohmic regime and is
strongly decreased for a critical value of the current.\cite{WAHL} One can
observe that the current value where the non-linearity sets in and the width
of the transition are strongly temperature dependent (both increase as the
temperature increases). Because of the rounding of the variation of $\frac R{%
Rohmic}$ $vs$ $I$, it is very difficult to define a value for a critical
current. However, a rough estimate gives $I_c$ = 4 mA and 1 mA for Pr$%
_{0.63} $Ca$_{0.37}$MnO$_3$ and Nd$_{0.6}$Ca$_{0.4}$MnO$_3$, respectively.
The temperature dependence of the lattice parameters in these compounds is
rather complex. It is well known that their temperature behavior is
characterized by the presence of changes of slopes corresponding to the
characteristic structural and magnetic ordering temperatures.\cite{COX98}
For T%
\mbox{$<$}%
$T_{CO}$, a comparison of the current {\it and} temperature dependence of
the cell volume determined from the synchrotron X-ray diffraction data
allows us to discard definitely any Joule heating effect. Indeed, no
evolution of the cell volume is observed as a function of the applied
current (Imax = 10 mA, well above the value of the current for which the
non-linearity sets in) whereas, in the range of temperature above 100K, a
steep increase of the cell volume is observed with zero biased current (See
figure 3 for Nd$_{0.6}$Ca$_{0.4}$MnO$_3$). Such a result is found for both
composition Pr$_{0.63}$Ca$_{0.37}$MnO$_3$ and Nd$_{0.6}$Ca$_{0.4}$MnO$_3.$

Figure 4a and 4b show for both compounds the integrated intensity of the
superlattice peaks corresponding to the CO as a function of the applied
current. In figure 5, a characteristic portion of the X-ray diffraction
patterns is shown at 100K with I =0, 1 and 5 mA.\ The shapes and the
intensities of the superlattice peaks are not modified whatever the value of
the applied current lying in the ohmic or non-ohmic region. Such a result
does not give a direct interpretation of the process involved in the
non-linear conduction but, at least, allows us to discard any CO
destabilization as an origin for this feature. Other interpretations, based
on a preserved superstructure when a current is flowing, must now be
proposed. In charge ordered compounds, observation of non-linear conduction
setting in along with a large broad band noise suggests that the charge
ordered state gets depinned at the onset of the non-linear regime.\cite
{GUH99,RAO00} Besides, the possibility of a charge density wave (CDW)
condensate in this charge ordered regime of the manganese oxides is now well
established.\cite{KIDA02,CHU01,ASAKA02} This might be a track to explain the
occurrence of the non-linear conduction in such systems. Recently, we have
proposed an interpretation in terms of collective excitation of CDW's.\cite
{WAHL} In such a framework, the non-linear conduction arises from the motion
of the CDW domains which carries a net electrical current.

\smallskip

\section{Figures Captions}

\begin{description}
\item[Figure 1]  : Portion of the synchrotron x-ray diffraction patterns for
Nd$_{0.6}$Ca$_{0.4}$MnO$_3$ at 100K with zero bias, in the region of the $%
(121)$ (a)$,$ $(302),$ (122) (b) and $(123)$ (c) superlattice peaks.

\item[Figure 2]  : $\frac R{R_{Ohmic}}$ versus bias current at 100K for Nd$%
_{0.6}$Ca$_{0.4}$MnO$_3$ (a) and Pr$_{0.63}$Ca$_{0.37}$MnO$_3$ (b). The
arrow denotes the estimated treshold current.

\item[Figure 3]  : Cell volume versus temperature for zero bias (filled
symbol, lower scale) and cell volume versus bias current at 100K (upper
scale)

\item[Figure 4]  : Integrated intensity of the $\sup $erlattice peaks as a
function of the bias current at 100K. a) Nd$_{0.6}$Ca$_{0.4}$MnO$_3$ ; b) Pr$%
_{0.63}$Ca$_{0.37}$MnO$_3.$ The arrows denote the occurence of the
non-linear regime.

\item[Figure 5]  Portion of the x-ray diffraction patterns for Nd$_{0.6}$Ca$%
_{0.4}$MnO$_3$ at 100K with bias 0 mA, 1 mA and 5 mA in the region of the $%
(123)$ superlattice peak.
\end{description}

\smallskip

\end{document}